\documentstyle[11pt,epsfig]{article}
\def\la{\mathrel{\mathchoice {\vcenter{\offinterlineskip\halign{\hfil
$\displaystyle##$\hfil\cr<\cr\sim\cr}}}
{\vcenter{\offinterlineskip\halign{\hfil$\textstyle##$\hfil\cr
<\cr\sim\cr}}}
{\vcenter{\offinterlineskip\halign{\hfil$\scriptstyle##$\hfil\cr
<\cr\sim\cr}}}
{\vcenter{\offinterlineskip\halign{\hfil$\scriptscriptstyle##$\hfil\cr
<\cr\sim\cr}}}}}
\title{{\bf Faint companions in the close environment of star--forming
dwarf galaxies: possible overlooked starburst triggers?
\footnote{Supported by German DFG grant FR325/50-1, IAC Summer
Research Programme 1998, Spanish DGES grant PB97-0158, and German DARA
GmbH grant 50 OR 9907 7.  We have made use of the NASA/IPAC
Extragalactic Database (NED) and the Lyon-Meudon Extragalactic
Database (LEDA).  We thank S.A. Pustilnik and A. Kniazev for fruitful
discussions and helpful comments.}}}
\author{K.G.~Noeske$^1$,
J.~Iglesias--P\'aramo$^2$, J.M.~V\'{\i}lchez$^3$, P.~Papaderos$^1$,
K.J.~Fricke$^1$\\
\vspace{0.1cm}\\
\normalsize $^1$Universit\"{a}ts-Sternwarte G\"{o}ttingen, D-37083 G\"ottingen, Germany\\
\normalsize $^2$IAC, 38200 La Laguna, Tenerife, Spain;
\normalsize $^3$IAA (CSIC), 18080 Granada, Spain
}
\date{}
\begin{document}
\maketitle
\def\bull{\vrule height .9ex width .8ex depth -.1ex}
\makeatletter
\def\ps@plain{\let\@mkboth\gobbletwo
\def\@oddhead{}\def\@oddfoot{\hfil\tiny
``Dwarf Galaxies and their Environment'';
Bad Honnef, Germany, 23-27 January 2001; Eds.{} K.S. de Boer, R.-J.Dettmar, U. Klein; Shaker Verlag}%
\def\@evenhead{}\let\@evenfoot\@oddfoot}
\makeatother

\begin{abstract}\noindent
Using the NASA Extragalactic Database, we have searched the close environment of
98 star--forming dwarf galaxies (SFDGs) from field-- and low density environments
for companion galaxies. Most of the found companions are dwarf galaxies, previously
disregarded in environmental studies of SFDGs.
Using a subsample at low redshifts, $cz <$ 2000 \,km\,s$^{-1}$, i.e. less biased
against dwarf companions, we find that 30\,\% of the SFDGs have close companions
within a projected linear separation $s_p <$100\,kpc and a
redshift difference of $\Delta cz <$ 500\,km\,s$^{-1}$.
This fraction must be considered a lower limit, given the incompleteness
of the available data sets and the non-negligible frequency of {H}{\small I}
clouds in the vicinity of SFDGs, so that the majority of SFDGs should not be
considered isolated.

The redshift differences between companion candidates and sample SFDGs are
typically $\la$ 250 km\,s$^{-1}$ and concentrated towards lower values. This is
similarly observed for dwarf satellites of spiral galaxies and suggests a physical
association between the companion candidates and the sample SFDGs.
SFDGs with a close companion do not show significant differences in their
H$\beta$ equivalent widths and $B-V$ colours as compared to isolated ones. However,
the available data do not allow to rule out close dwarf companions as an influencing
factor for star formation activity.
\end{abstract}
%
%
\section{Introduction}
%
%
\begin{figure}[!t]
\centerline{
\raisebox{0.5mm}{\epsfig{figure=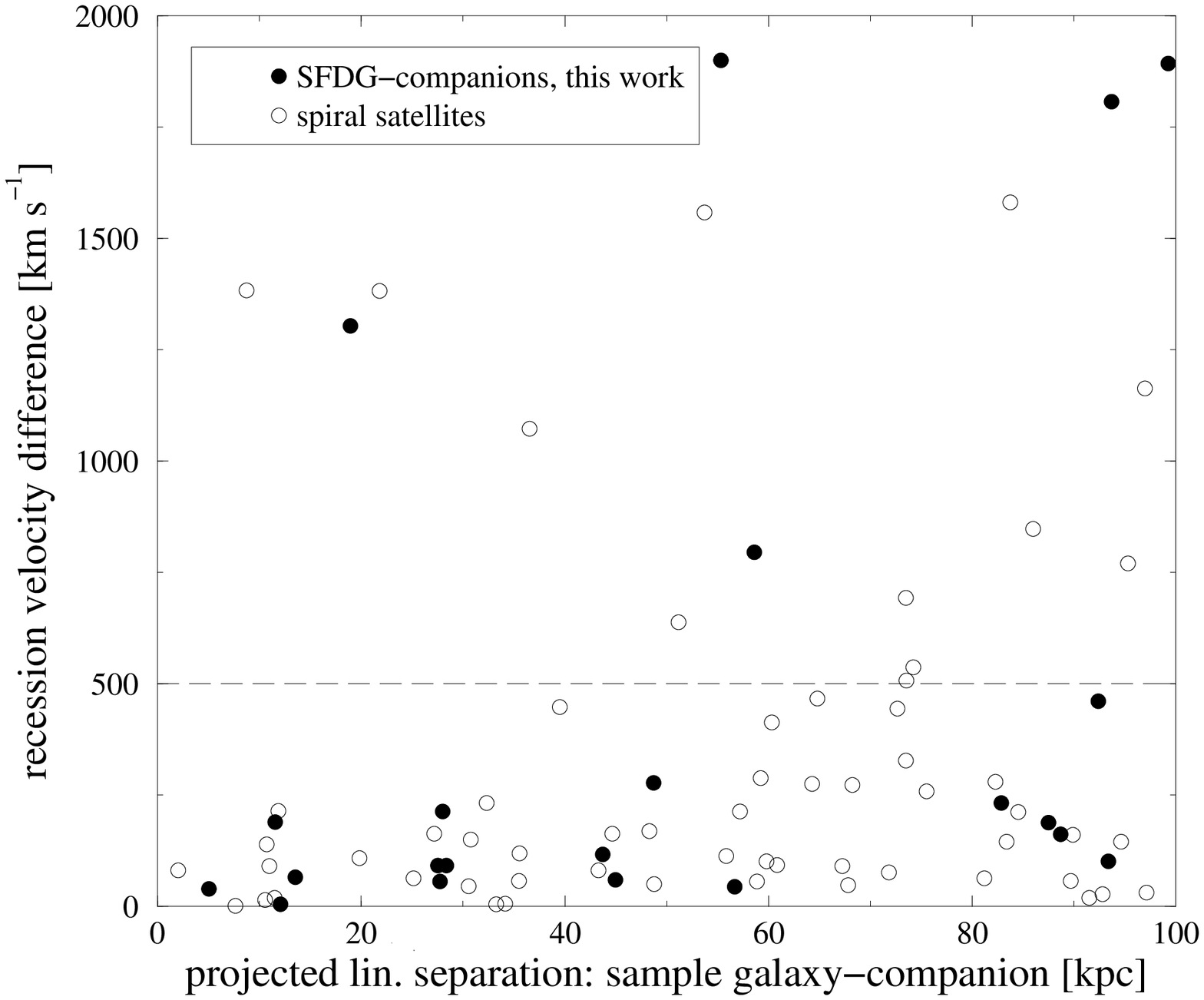,height=5.05cm}\hspace*{1cm}}
\epsfig{figure=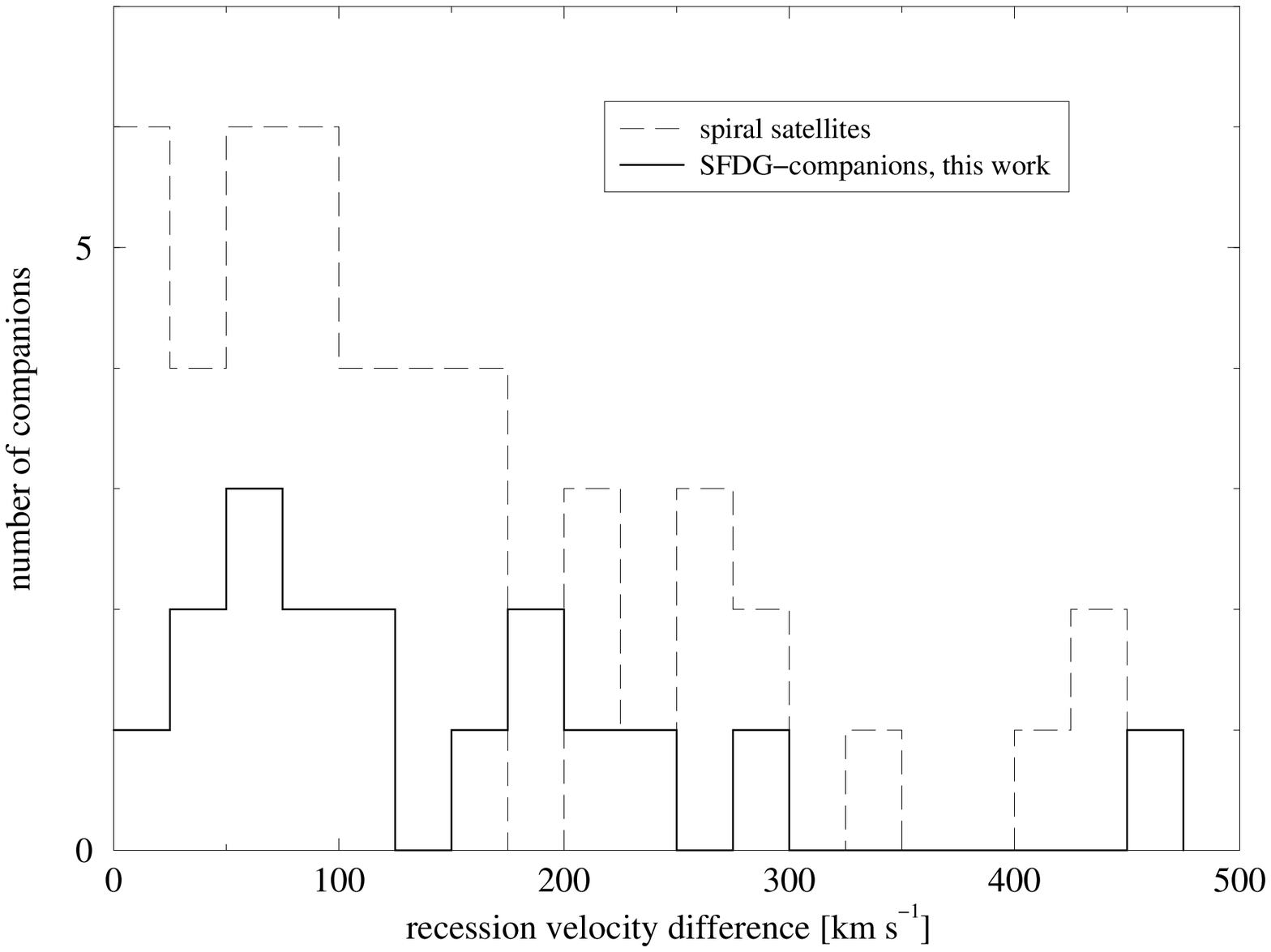,height=5cm}
}\ \\[-2em]
\caption{\small {\bf left} Recession velocity difference vs. projected linear
separation of putative companions of SFDGs (filled symbols).
Open circles show the distribution of companions (not restricted to dwarfs) we
found around field spiral galaxies from the sample of Kennicutt \& Kent (1983).
{\bf right} Distribution of the recession velocity differences between the sample
SFDGs and the putative companions. The dashed histogram represents the spiral
companions shown in the left panel of this figure.}
\label{com_dv}
\end{figure}
%
%
Comparative studies of gas--rich dwarf galaxies with current or recent strong star
formation (SF) suggest that they form one and the same physical class of objects
(Papaderos et al. 1996, Marlowe et al. 1999). In the following, we shall therefore
unify Blue Compact Dwarf Galaxies (BCDs), H{\small II} galaxies, etc. using
the term ``Star--Forming Dwarf Galaxies (SFDGs)''
(cf. V\'{\i}lchez 1995).
Starbursts or episodes of strongly enhanced star formation (SF) are considered frequent
in gas-rich dwarf galaxies, and are principal evolutionary drivers of such objects.
The frequency of local dwarfs implies numerous less evolved, i.e. gas--rich,
probably starbursting dwarfs at higher redshifts, possibly contributing significantly
to the star formation rate (SFR) density (see e.g. Guzm\'an et al. 1997).
While internal regulation processes were put forward to explain
their strong SF activity (cf. Papaderos et al. 1996 and references therein), SFDGs were
also suggested to be influenced by interaction with their environment.
Studies assessing this hypothesis revealed that SFDGs typically reside in regions of lower
density of luminous galaxies (Salzer 1989, Telles \& Terlevich 1995).
Interactions with luminous galaxies, given their typically large distances, were therefore
not considered a general trigger mechanism for starburst activity in SFDGs. This scenario
was also abandoned in view of comparisons between SFDGs with and without  --- generally
relatively distant --- luminous companions
(Campos--Aguilar \& Moles 1991, Telles \& Terlevich 1995).
On the other hand, mixed evidence for differing properties of SFDGs
in void, field and cluster environments (V\'{\i}lchez 1995, Vennik et al. 2000),
although controversially discussed, calls for an assessment of alternative
environmental factors.
In this respect, it appears interesting that Lindner et al. (1996) attributed the
apparent extreme isolation of some distant BCDs to the observational bias against
faint companions.
A significant influence of a low mass companion close to a SFDG appears possible,
since the tidal forces it exerts scale only with the first power of its mass, but
with the third power of its distance. An alternative scenario is the triggering and
fueling of a starburst by infall of gas--rich companions onto a SFDG (e.g. Hensler et al.
1999). Observational results suggest that such yet uncatalogued, optically faint
objects, down to extragalactic H{\small I} clouds with no optical counterpart, are
frequent around SFDGs (e.g. Taylor et al. 1995, Pustilnik et al. 1997).
We present a first study of the frequency, properties and possible influence
of such companions in the close environment of SFDGs.
For a detailed description of this work, see Noeske et al. (2001).
\begin{figure}[!t]
\hspace{0.5cm}
\centerline{
\epsfig{figure=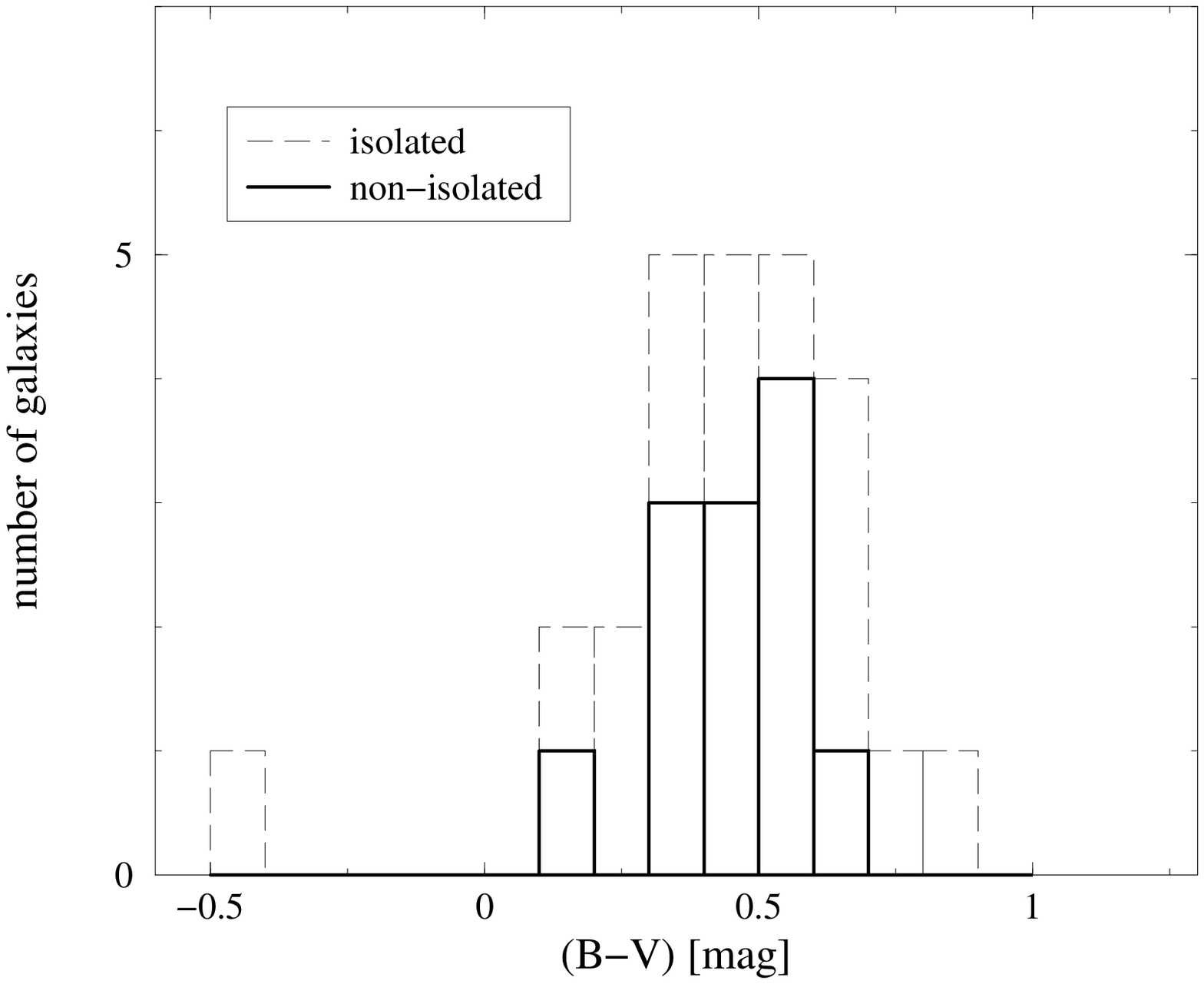,height=5cm}\hspace{1cm}
\raisebox{0.5mm}{\epsfig{figure=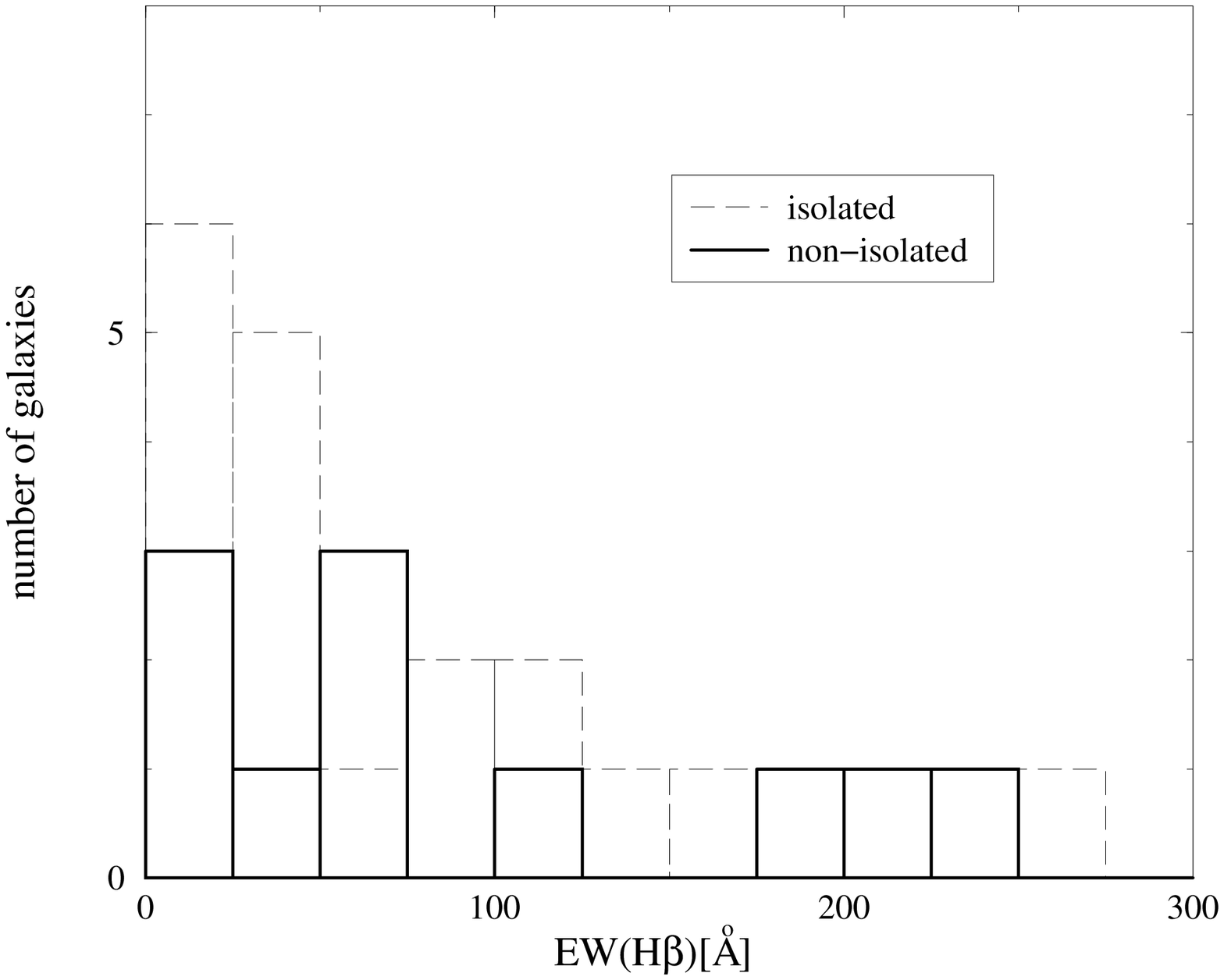,height=5cm}}
}\ \\[-2em]
\caption{\small Distribution of the {\bf left:} $B - V$ colours and
{\bf right:} H$\beta$ equivalent widths $EW($H$\beta )$ for the
isolated (dashed line) and the non--isolated (solid line) subsample of
SFDGs. The comparison is restricted to the subsample at close
distances, i.e. at redshifts $cz <$\,2000\,km\,s$^{-1}$.}
\label{histo}
\end{figure}
%
%
\begin{table}[ht]
\begin{small}
\caption{\small\noindent Companion search results for subsamples within different redshift
intervals}
\label{biastab}
\begin{center}
\begin{tabular}{lccccc}
\hline
$cz^a$\,[km\,s$^{-1}$] & sample & non&  no. of & dwarf \\
range & size & isol. & comps.&  comps. \\
(1)     & (2)   & (3)   & (4)   & (5) \\
\hline
unconstrained            & 98  & 16 & 18 & 15 \\
$cz < 2000$           & 42   & 13 & 15 & 14 \\
$2000 \leq cz < 4000$\hspace{0.5cm} & 12   & 2  & 2  & 1 \\
$cz \geq 4000$         & 44   & 1  & 1  & 0 \\
\hline
\end{tabular}
\end{center}
$^a$ redshifts correspond to Virgo infall corrected distances,
assuming $H_0$\,=\,75\,km\,s$^{-1}$\,Mpc$^{-1}$;
(1) redshift interval of respective subsample;
(2) no. of sample galaxies in subsample;
(3) no. of sample galaxies in subsample with at least one possible
companion;
(4) no. of possible companions found for the respective subsample;
(5) no. of dwarf galaxies among possible companions
\end{small}
\end{table}
\section{The SFDG sample and companion search catalog}
%
We selected all dwarf H{\small II} galaxies from the University of Michigan
Lists IV and V, studied in detail by Salzer et al. (1989). As these objects
are relatively distant (mostly $cz\,>\,$2000\,km\,s$^{-1}$), we included a number
of well--observed BCDs at smaller distances (Cair\'os 2000) to be
less biased against low--luminosity companions.
None of our objects resides in a group or cluster environment.
To search for companion objects, we chose the NASA Extragalactic Database (NED),
one of the deepest online catalogs to date. Its inhomogeneous completeness does
not hamper our study, as one still cannot extrapolate even beyond a sharp limit
 (cf. Section \ref{comp_bias}).
We searched for extragalactic optical sources with the additional requirement of
a known redshift, so that the companions' properties and distribution could be studied.
Spectrophotometric data from the NED and different literature sources were compiled
for both the sample SFDGs and the found companions. Distances were calculated from
NED redshifts, using Tully (1988) to correct for Virgo Cluster infall.

\section{Results and Discussion}
\subsection{Companion selection criteria and distribution of the companions}
%
Figure \ref{com_dv} (left) shows that almost all of the companions have
low redshift differences ($\la\,$500\,km\,s$^{-1}$) to the sample SFDGs.
Subsequently, a projected linear separation $s_p <$\,100\,kpc and a
redshift difference corresponding to $\Delta cz <$\,500\,km\,s$^{-1}$ were adopted
as companion selection criteria, values below which tidal forces have been estimated
to be significant (Campos--Aguilar et al. 1993, Pustilnik et al. 2001) and for which
pairs of normal galaxies and SFDGs show likely signs of interaction.
The rising frequency of companions towards lower redshift differences
(Figure \ref{com_dv}, right) is reminiscent of what is observed for binary galaxies
(Schneider \& Salpeter 1992) as well as for dwarf companions of normal
spiral galaxies (Zaritsky et al. 1997; cf. also Figure \ref{com_dv} left \& right for
spiral companions of any luminosity).
It is hence very likely that the companions we find are physically associated
with the sample SFDGs, rather than random encounters.
The independence of $\Delta cz$ on $s_p$ found for spiral companions has been
attributed to the dynamic dominance of a massive DM halo (Zaritsky et al. 1997).
A similar scenario appears tempting for SFDG companions, if a larger dataset can prove
that they also show no correlation of $s_p$ and $\Delta cz$, as suggested by Figure
\ref{com_dv} (left).

%
\subsection{Companion properties: invisible distant dwarfs}
\label{comp_bias}
%
Table \ref{biastab} (first row) shows that the majority ($>$80\%) of the found
companions are dwarf galaxies (M$_B >$\,--18\,mag). As expected, these are
almost solely found for the nearest subsample ($cz < 2000$\,km\,s$^{-1}$) due to their
intrinsic faintness. From this closest, i.e. least biased subsample, one obtains a
fraction of $\sim$30\% of SFDGs with at least one close companion. We emphasize that
this is a lower limit, given the incompleteness of our search catalog. Unfortunately,
as both the faint end of the galaxy luminosity function, and the frequency of purely
gaseous companions are still poorly constrained, a meaningful extrapolation below this
limit does not appear reasonable to date.
The average companion is as blue as the sample SFDGs ($\overline{B-V}\,=$\,0.44\,mag),
and by 0.72\,B\,mag (median) brighter than its 'mother' galaxy. This probably reflects a
selection effect, as preferably the brightest companions -- nevertheless dwarfs --
with active SF are catalogued in the NED.

%
\subsection{Close dwarf companions as possible starburst triggers?}
%
Objects with and without found companions ('isolated' and 'non--isolated') are partly
separable only for the nearest ($cz < $\,2000\,km\,s$^{-1}$) subsample. These galaxies
were compared with respect to $B-V$ and $EW($H$\beta )$, observables which are available
for most of the sample and trace the relative SF activity on respective timescales
of $\sim\,$10$^7$ and $\sim\,$10$^8$\,yr. The distributions for the isolated and
non--isolated SFDGs of the nearest subsample are shown in Figure \ref{histo}.
Kolmogorov--Smirnov tests yield high probabilities for equal parent distributions
($B-V$: 0.90, $EW($H$\beta )$: 0.45), whereas the respective sample means
are compatible within the sample standard deviations.
These results can neither prove the hypothesis of purely internal triggering of SF,
nor disprove an external influence of close companions, due to the low number
statistics and the large intrinsic scatter of the data.
The picture is further blurred by the incompleteness of the companion search, i.e.
a poor separation of isolated and non--isolated sample objects. In addition, the time
window to detect a starburst through {\em strongly} changed SF tracers is narrow for
SFDGs, $\sim$10$^7$\,yr after its onset.
The presence of dwarf companions in the close environment of SFDGs readdresses the
question of environmental influences on their SF activity, a scenario which had been
abandoned before due to the general lack of luminous companions. To obtain decisive
answers, large samples and deep search catalogs will be required, as well as advances
on the part of theory on interacting dwarf galaxies.

%
{\small
\begin{description}{} \itemsep=0pt \parsep=0pt \parskip=0pt \labelsep=0pt
\item {\bf References}

\item[]Cair\'os, L.M. 2000, PhD Thesis, Univ. de La Laguna

\item[]Campos-Aguilar, A., Moles, M. 1991, A\&A, 241, 358

\item[]Campos-Aguilar, A., Moles, M., Masegosa, J. 1993, AJ, 106, 1784

\item[]Guzm\'an, R., et al. 1997, ApJ, 489, 559

\item[]Hensler, G., Rieschick, A., K\"oppen, J. 1999, Ap\&SS,
in press (astro-ph/9908242)

\item[]Kennicutt, R.C., Kent, S.M. 1983, AJ, 88, 1094

\item[]Lindner, U., et al. 1996, A\&A, 314, 1

\item[]Marlowe, A.T., Meurer, G.R., Heckman, T.M. 1999, ApJ, 522, 183

\item[]Noeske, K.G., Iglesias--P\'aramo, J., V\'{\i}lchez, J.M., Papaderos, P.,
Fricke, K.J. 2001, A\&A, in press

\item[]Papaderos, P., Loose, H.-H., Fricke, K.J.,
Thuan, T.X. 1996, A\&A, 314, 59

\item[]Pustilnik, S.A., Kniazev, A.Y., Ugryumov, A.V. 1997, IAUJD, 2E, 60


\item[]Pustilnik, S.A., Kniazev, A.Y.,
Lipovetsky, V.A., Ugryumov, A.V. 2001, A\&A, in press

\item[]Salzer, J.J. 1989, ApJ, 347, 152

\item[]Salzer, J.J., McAlpine, G.M., Boroson, T.A. 1989, ApJS, 70, 447


\item[]Schneider, S.E., Salpeter, E.E. 1992, ApJ, 385, 32

\item[]Taylor, C.L., Brinks, E., Grashius, R.M., Skillman, E.D., 1995,
ApJS 102, 189

\item[]Tully, R.B. 1988, Nearby Galaxies Catalog, Cambridge University Press

\item[]Vennik, J., Hopp, U., Popescu, C.C. 2000, A\&AS, 142, 399

\item[]V\'{\i}lchez, J.M. 1995, AJ, 110, 1090

\item[]Zaritsky, D., Smith, R., Frenk, C., White, S.D.M. 1997, ApJ, 478, 39

\end{description}
}

\end{document}